\documentclass[runningheads]{llncs}

\usepackage{times}
\usepackage{amsfonts} 
\usepackage{amsmath} 
\usepackage{amssymb} 
\usepackage[all]{xy} 

\begin{document}

\title{On the {S}ecurity of {W}ang's {P}rovably {S}ecure {I}dentity-based {K}ey {A}greement {P}rotocol}

\author{Maurizio Adriano Strangio}
\institute{University of Rome ``Roma Tre", ROME, ITALY\\
\email{strangio@mat.uniroma3.it}\\ }

\newcommand{\fracnoline}[2]{\begin{array}{c}{#1}\\{#2}\end{array}}
\newcommand{\assign}[2]{{#1}\stackrel{}{\leftarrow}{#2}}
\newcommand{\passign}[2]{{#1}\stackrel{R}{\leftarrow}{#2}}
\newcommand{\dpassign}[3]{{#1}\stackrel{#2}{\leftarrow}{#3}}
\newcommand{\tabbegin}[0]{\begin{tabular}{p{0.5cm}p{1.5cm}p{9.5cm}}}
\newcommand{\tabend}[0]{\end{tabular}}
\newcommand{\fracnl}[2]{\begin{center}\begin{tabular}{c}{#1}\\{#2}\end{tabular}\end{center}}

\newtheorem{Def}{Definition}
\newtheorem{Th}{Theorem} 
\newtheorem{Assump}{Assumption}

\maketitle

\begin{abstract}
In a 2005 IACR report, Wang published an efficient identity-based key agreement protocol (IDAK) suitable for resource constrained devices. 

The author shows that the IDAK key agreement protocol is secure in the Bellare-Rogaway model with random oracles and also provides separate ad-hoc security proofs claiming that the IDAK protocol is not vulnerable to Key Compromise Impersonation attacks and also enjoys Perfect Forward Secrecy (PFS).

In this report, we review the security properties of the protocol and point out that it is vulnerable to Unknown Key Share attacks. Although such attacks are often difficult to setup in a real world environment they are nevertheless interesting from a theoretical point of view so we provide a version of the protocol that fixes the problem in a standard way. We also provide a security proof of the IDAK protocol based on the Gap Bilinear Diffie Hellman and random oracle assumptions  in the stronger extended Canetti-Krawczyk security model of distributed computing. 

\end{abstract}

\section{Introduction}
\label{sec:INTRO}

In a 2005 IACR report \cite{WANG05-IACR} and more recently in the arXiv archive \cite{WANG12}, Wang proposed a novel identity-based key agreement protocol (IDAK) using the Weil/Tate pairing and also provided a security proof in the Bellare-Rogaway model \cite{BR93}. 
The model does not capture Key Compromise Impersonation (KCI) attacks or Perfect Forward Secrecy (PFS); for the later security features the author provides separate ad-hoc proofs. 

However, the protocol is vulnerable to Unknown Key Share (UKS) attacks. Although such attacks are often difficult to setup in a real world environment they are nevertheless interesting from a theoretical point of view so we provide a version of the protocol that fixes the problem in a standard way.

We also show that the IDAK protocol is provably secure in the extended Canetti-Krawczyck (eCK) model \cite{MLM07} under the Gap Bilinear Diffie Hellman and random oracle assumptions.

This paper supercedes a report withdrawn from the IACR Cryptology ePrint Archive where the author presented an erroneous result on Wang's protocol \cite{IACR13}.

\section{Notation and mathematical background}
\label{sec:MATH}
To make the paper self-contained, we briefly recall the underlying mathematical concepts and notation. Let us consider two multiplicative cyclic groups $G$ and $G_1$ of order $q$ with $g$ a generator of $G$. The bilinear map $\hat{e}:G\times G \rightarrow G_1$ has the following three properties:
\begin{enumerate}
	\item bilinearity, for all $g_1,g_2\in G$ and $x,y \in Z: \hat{e}(g_1^x,g_2^y)=\hat{e}(g_1,g_2)^{xy}=\hat{e}(g_1^y,g_2^x)$;
	\item non-degeneracy, for all $g \in G$, $\hat{e}(g,g)\not=1$ is a generator in $G_1$;
	\item computability, for $g_1, g_2 \in G: \hat{e}(g_1,g_2) \in G_1$ is computable in polynomial time. 
\end{enumerate}
 The modified Weil and Tate pairings associated with supersingular elliptic curves are examples of admissible pairings \cite{CHENKUDLA02-IACR}, \cite{BF01}.

If $X$ is a finite set then $\passign{x}{X}$ or $x \in_R X$ denote the sampling of an element uniformly at random from $X$. If $\alpha$ is neither an algorithm nor a set $\assign{x}{\alpha}$ represents a simple assignment statement. 

The (computational) Bilinear Diffie-Hellman assumption (BDH) holds in the group $G$ if for random elements $x,y,z \in Z_q^*$ it is computationally hard to compute $\hat{e}(g,g)^{xyz}$.

\begin{Assump}[BDH] The group G satisfies the Bilinear Diffie-Hellman assumption if for all PPT algorithms we have: 
\begin{displaymath}
\begin{array}{l}
\passign{x}{Z_q^*}; \passign{y}{Z_q^*}; \passign{z}{Z_q^*};
\assign{X}{g^x};\assign{Y}{g^y};\assign{Z}{g^z}: \\
\emph{Pr}\left[\mathcal{A}(X,Y,Z)\emph{=}\hat{e}(g,g)^{xyz}\right] < \epsilon \\
\end{array}
\end{displaymath}
where the probability is taken over the coin tosses of $\mathcal{A}$ (and random choices of $x,y,z$) and $\epsilon$ is a negligible function.
\label{assump-CDHP}
\end{Assump}

The Decisional Bilinear Diffie-Hellman assumption (DBDH) holds in the group $G$ if for random elements $x,y,z,r \in Z_q^*$ it is computationally hard to distinguish the distributions $\langle g^x,g^y,g^z,g^r \rangle$ and $\langle g^x,g^y,g^z,\hat{e}(g,g)^{xyz} \rangle$.

\begin{Assump}[DBDH] The group G satisfies the Decisional Bilinear Diffie-Hellman Assumption if for all PPT algorithms we have: 
\begin{displaymath}
\begin{array}{l}
\passign{x}{Z_q^*}; \passign{y}{Z_q^*}; \passign{z}{Z_q^*}; \passign{r}{G};
\assign{X}{g^x};\assign{Y}{g^y};\assign{Z}{g^z}: \\
\emph{Pr}[\mathcal{A}(X,Y,Z,r)\emph{=}1]\emph{-}\emph{Pr}[\mathcal{A}(X,Y,Z,\hat{e}(g,g)^{xyz})\emph{=}1] < \epsilon
\end{array}
\end{displaymath}
where the probability is taken over the coin tosses of $\mathcal{A}$ (and random choices of $x,y,z,r$) and $\epsilon$ is a negligible function.
\label{assump-DBDH}
\end{Assump}

The Gap Bilinear Diffie-Hellman assumption (GBDH) holds in the group $G$ if for random elements $x,y,z \in Z_q^*$ it is computationally hard to solve the BDH problem even with access to a DBDH oracle.

\begin{Assump}[GBDH] The group G satisfies the Gap Bilinear Diffie-Hellman Assumption if for all PPT algorithms we have: 
\begin{displaymath}
\begin{array}{l}
\passign{x}{Z_q^*}; \passign{y}{Z_q^*}; \passign{z}{Z_q^*}; 
\assign{X}{g^x};\assign{Y}{g^y};\assign{Z}{g^z}: \\
\assign{\emph{DBDH}(g,g^a,g^b,g^c,h)}{1\emph{ iif }h=e(g,g)^{abc}};\passign{a}{Z_q^*}; \passign{b}{Z_q^*}; \passign{c}{Z_q^*};\\
\emph{Pr}\left[\mathcal{A}(X,Y,Z)\emph{=}\hat{e}(g,g)^{xyz}\right] < \epsilon \\
\end{array}
\end{displaymath}
where the probability is taken over the coin tosses of $\mathcal{A}$ (and random choices of $x,y,z$) and $\epsilon$ is a negligible function.
\label{assump-GBDH}
\end{Assump}

\section{Review of the IDAK protocol}
\label{sec:IDAK} 

In this section we review the IDAK identity-based key agreement protocol.
The protocol is completely specified by three algorithms Setup, Extract, Exchange \cite{WANG05-IACR}:
\begin{itemize}
\item \textbf{Setup}, for input the security parameter $k$:
\begin{enumerate}
	\item Generate a bilinear group $G_{\rho} = \{G,G_1, \hat{e}\}$ with the groups $G$ and $G_1$ of prime order $q$. Define $h$ as the co-factor of the group order q for G; 
	\item Choose a generator $g \in G$;
	\item Choose a random master secret key $\alpha \in_R Z_q^*$;
	\item Choose the cryptographic hash functions $H: \{0, 1\}^* \rightarrow G$ and $\pi: G \times G  \rightarrow Z_q^*$. In the security analysis of protocol IDAK, $H$ and $\pi$ are simulated as random oracles.
\end{enumerate}
The system parameters are $(hq, h, g, G, G_1, \hat{e}, H, \pi)$ and the master secret key is $\alpha$.

\item \textbf{Extract}, For a given identification string $ID \in \{0, 1\}^*$, the algorithm computes $g_{ID} = H(ID) \in G$ and returns the private key $d_{ID} = g_{ID}^{\alpha}$;

\item \textbf{Exchange}, For two peers $A$ and $B$ with identities ID$_A$ and ID$_B$ respectively, the algorithm proceeds as follows:
\begin{enumerate}
	\item $A$ selects $x \in_R Z_q^*$, computes $R_A = g_{ID_A}^x$ and sends $R_A$ to $B$;
	\item $B$ selects $y \in_R Z_q^*$, computes $R_B = g_{ID_B}^y$ and sends $R_B$ to $A$;
	\item On receipt of $R_B$, $A$ computes $s_A = \pi(R_A,R_B), s_B = \pi(R_B,R_A)$ and the shared secret $sk_{AB}$ as \\$\hat{e}(g_{ID_A}, g_{ID_B})^{(x+s_A)(y+s_B)h\alpha} = \hat{e}(g_{ID_B}^{s_B}\cdot R_B, g_{ID_A}^{(x+s_A)h\alpha})$;
	\item On receipt of $R_A$, $B$ computes $s_A = \pi(R_A,R_B), s_B = \pi(R_B,R_A)$ and the shared secret $sk_{BA}$ as \\$\hat{e}(g_{ID_A}, g_{ID_B})^{(x+s_A)(y+s_B)h\alpha} = \hat{e}(g_{ID_A}^{s_A}\cdot R_A, g_{ID_B}^{(x+s_B)h\alpha})$;
\end{enumerate}
\end{itemize}

The main result of \cite{WANG05-IACR} is Theorem 5.2 which proves that IDAK is a secure key agreement protocol in the Bellare-Rogaway model under the DBDH and random oracle assumptions. The author also presents ad ad-hoc security proofs claiming that the protocol enjoys PFS (Theorem 6.1) and is not vulnerable to KCI attacks (Theorem 7.1).

\section{The security features of the IDAK protocol}
\label{sec:IDAKSEC} 

In this section we review the security properties of the IDAK protocol.

\subsection{Entity Impersonation}

When an honest party $A$ is corrupted then a malicious adversary can easily impersonate him in a protocol execution with another party $B$ since the adversary has total control of any session initiated with identity $A$. 
The adversary can also register a new identity (say $C$) with the KGC and successfully conduct an impersonation attack against $B$.
Under such circumstances, no ``implicitly authenticated" key agreement protocol can resist against entity impersonation attacks.

\subsection{Forward Secrecy}
\label{sec:IDAKFS}

A key agreement protocol has perfect forward secrecy (PFS), if after a communication  session between two peers is completed, the adversary cannot learn the session key even if it corrupts both parties involved in that session. In other words, learning the long term private keying material of parties should not compromise the security of past completed sessions.

A weaker notion of forward secrecy (weak perfect forward secrecy $-$ wPFS) introduced by Krawczyk \cite{HK05} considers only completed sessions in which the adversary was passive (i.e. the later did not modify the messages transcripts exchanged by the parties). The IDAK protocol is in fact wPFS-secure. 

For identity protocols ``master key PFS" (mkPFS) refers to the consequences of KGC corruption. The IDAK protocol does not possess mkPFS since knowledge of the secret master key $\alpha$ would allow even a passive adversary to trivially break the protocol since the session key of two peers $A,B$ is calculated as $\hat{e}(g_{ID_B}^{s_B}\cdot R_B, R_A \cdot g_{ID_A}^{s_A})^{\alpha}$. 

To make the IDAK protocol mkPFS Wang proposes the exchange of a an additional Diffie-Hellmann ephemeral key that should be included in the computation of the session key (see Section 6 in \cite{WANG05-IACR}).

\subsection{Key Compromise Impersonation}
\label{sec:IDAKKCI}

In a KCI attack, the opponent has learned the long-term private key of an honest party (say $A$) and attempts to establish a valid session key with $A$ by masquerading as another legitimate principal (say $B$). 
This attack represents a subtle threat that is often underestimated and difficult to counter \cite{KCISTR06-IACR}. 

Suppose a malicious adversary $\mathcal{A}$ has learned the long term private key
$d_{ID_A}$ of principal $A$; she is now able to set up a man-in-the-middle attack during a run of the protocol 
between $A$ and $B$. The attack should work as follows. $\mathcal{A}$ lets message $R_A$
reach its intended destination ($B$) but replaces $B$'s response $R_B$ to $A$ with a transcript $X$. 
On receipt of $X$, $A$ calculates its session key as follows:

\begin{eqnarray*} 
sk_{AB} & = & \hat{e}(g_{ID_B}^{s_B}\cdot X, d_{ID_A}^{(x+s_A)}) \\ 
& = & \hat{e}(g_{ID_B}^{s_B}\cdot X, (g_{ID_A}^{x} \cdot  g_{ID_A}^{s_A})^{h\alpha}) \\
& = & \hat{e}(d_{ID_B}^{s_B}\cdot X^{\alpha}, R_{A} \cdot  g_{ID_A}^{s_A}) \\
\end{eqnarray*}  

To succeed in the attack $\mathcal{A}$ must choose a suitable message $X$ possibly exploiting the algebraic properties of the underlying groups in order to have $A$ accept  
a known session key. To eliminate the dependency from $\alpha$ the only possible choice is to set $X=g_{ID_B}$ so that $X^{\alpha}=d_{ID_B}$, however, the adversary would have to corrupt $B$ also to succeed.

A more powerful adversary, with the possibility of corrupting $B$'s session state in order to obtain the value of $y$, may succeed in breaking the protocol. The scenario wherein the adversary has additional information is not technically a KCI attack; such capabilities are considered in the security models of Canetti and Krawczyk \cite{CK01} and Lamacchia \emph{et al.} \cite{MLM07}. 

\subsection{Unknown Key Share}
\label{sec:IDAKUKS}

The first Unknown Key Share (UKS) attack against a key agreement protocol was described by Diffie \emph{et al.} \cite{DOM92} who showed how a dishonest entity $E$ can setup an attack whereby an honest entity $A$ finishes a protocol execution  believing she shares a key with $B$, but $B$ mistakenly believes the key is shared with $E$ (while the later does not necessarily know the session key held by  $B$).

It is possible to mount a UKS attack against the IDAK protocol:

\begin{enumerate}
 \item Adversary $\mathcal{A}$ register identity $E$ with the KGC and obtains the private key $d_{ID_E}$;
  \item After $A$ negotiating a communication session with $B$, she sends message $R_A$ to $B$;
  \item $\mathcal{A}$  intercepts message $R_A$, and initiates a new communication session with $B$ as the legitimate peer $E$ by sending $R_E=g_{ID_E}^t$ for  $t \in_R Z_q^*$, thus replacing message $R_A$; 
 \item  On receipt of $R_E$, $B$ cannot detect the difference between $R_E$ and $R_A$ because they are indistinguishable and responds with message $R_B$;
 \item  $\mathcal{A}$ relays $R_B$ to $A$. As the result, $A$ accepts believing that she is communicating with $B$ (and indeed they calculate the same session key) while the latter peer also accepts but is convinced she is connected to $E$. In this case the adversary cannot compute the  key shared by $A,B$ but nevertheless the attack is successful.
\end{enumerate}

In most cases countermeasures are easily implemented to avoid this type of attacks, so it is not often a major concern in practice. In particular, a common solution is to add the peer identities in the calculation of the key as illustrated in Figure \ref{FIG-PROT-IDAK}.

\begin{figure*}[ht]
\centering
\tabbegin
  \hline
   & {$A:$}\raggedleft & $\passign{x}{Z_q^*}$ \\
   & & $\assign{R_A}{g_{ID_A}^x}$ \\
   & {$A \rightarrow B:$}\raggedleft & $R_A$ \\
   & {$B:$}\raggedleft & $\passign{y}{Z_q^*}$ \\
   & & $\assign{R_B}{g_{ID_B}^y}$ \\
   & {$B \rightarrow A:$}\raggedleft & $R_B$ \\
   & {$A:$}\raggedleft & $\assign{s_A}{ \pi(A,B,R_A,R_B)}$, $\assign{s_B}{\pi(B,A,R_B,R_A)}$ \\
   & & $\assign{\sigma}{ \hat{e}(g_{ID_B}^{s_B}\cdot R_B, d_{ID_A}^{(x+s_A)})}$ \\
   & & $\assign{sk_{AB}}{H_1(A,B,R_A,R_B,\sigma})$ \\
   & {$B:$}\raggedleft & $\assign{s_A}{ \pi(A,B,R_A,R_B)}$, $\assign{s_B}{\pi(B,A,R_B,R_A)}$ \\
   & & $\assign{\sigma}{ \hat{e}(g_{ID_A}^{s_A}\cdot R_A, d_{ID_B}^{(y+s_B)})}$ \\
   & & $\assign{sk_{BA}}{H_1(A,B,R_A,R_B,\sigma})$ \\
  \hline
\tabend
\caption {Protocol IDAK}
\label{FIG-PROT-IDAK}
\end{figure*}

\section{A formal security proof of the IDAK protocol in the Extended Canetti-Krawczyk model}
\label{sec:SECPROOF}

In this section we review the extended Canetti-Krawczyk (eCK) model of distributed computing of Lamacchia \emph{et al}. \cite{MLM07} and then prove that the IDAK protocol is eCK-secure. 

\subsection{The eCK model}
\label{SEC:ECK-MODEL} 
Parties and the adversary are modeled as probabilistic Turing machines. The network is totally under the control of the adversary who can delete, modify or inject messages that are exchanged by any two honest parties running the protocol. \\
\textbf{Sessions}. Instances of a protocol run by two any parties $A$ and $B$ are called \emph{sessions}. A Session Identifier (SID) consists of the identities of the two participants and the messages transcripts they exchanged during the communication. 
For any two parties $A$ and $B$, we denote by $sid$ the SID of the session owned by $A$ and by $sid^*$ the SID of the session owned by its intended peer $B$.\\
\textbf{Adversary}. To capture all types of attacks resulting from ephemeral and long-term data compromise  the adversary is allowed to ask the following queries
\begin{enumerate}
\item[a)] \texttt{EphemeralKeyReveal}($sid$) to obtain \emph{all} short-term secret information used by a party in session $sid$; 
\item[b)] \texttt{PrivateKeyReveal}($ID$) which returns long-term private keys of the principal with identity $ID$; 
\item[c)] \texttt{SessionKeyReveal}($sid$) for exposing the session key of session $sid$; 
\item[d)] \texttt{Extract}($ID$) allows the adversary to register a new identity $ID$ and receive the private keying material relative to this identity. 
\end{enumerate}


In the eCK security experiment the adversary can ask two additional queries
\begin{enumerate}
\item[a)] \texttt{Test}($sid$) when the adversary asks this query (only once) a random bit $b$ is used to decide whether to return the real session key of $sid$ or a random key; 
\item[c)] \texttt{Guess}($b'$) when the  adversary terminates it outputs bit $b'$ as its guess of $b$, the query returns 1 if the guess is correct 0 otherwise.
\end{enumerate}
Test sessions are classified as either ``passive'' or ``active''. In a passive session the adversary is limited only to observing the communication transcripts exchanged by two honest participants while in active sessions the adversary can also tamper with them (e.g. cancel or modify communication messages). As the result, in passive sessions two parties may complete matching sessions (i.e. the conversation of the initiator matches the conversation of the responder and both have accepted \cite{BR93}); on the other hand, active sessions are those where matching sessions are not necessarily established. 
The adversary succeeds in the eCK experiment if the selected test session is \emph{fresh} and \texttt{Guess} returns 1. 
Set out below is the definition of a \emph{fresh} session.

\begin{Def}[fresh session] Let $sid$ denote the SID of a completed session run by party $A$ with a peer $B$. Let $sid^{*}$ denote the SID of the matching session of $sid$ owned by $B$, if it exists. Session $sid$ is \emph{fresh} if none of the following conditions hold:
\begin{enumerate}
\item the adversary issues a \emph{\texttt{SessionKeyReveal}}($sid$) query or\\ a \emph{\texttt{SessionKeyReveal}}($sid^{*}$) query (if $sid^{*}$ exists);
\item $sid^{*}$ exists and the adversary makes either of the following queries: \\(a)  \emph{\texttt{PrivateKeyReveal}}($A$) and \emph{\texttt{EphemeralKeyReveal}}($sid$) or 
\\(b)  \emph{\texttt{PrivateKeyReveal}}($B$) and \emph{\texttt{EphemeralKeyReveal}}($sid^{*}$);
\item $sid^{*}$ does not exist and the adversary makes either of the following queries: \\(a) \emph{\texttt{PrivateKeyReveal}}($A$) and \emph{\texttt{EphemeralKeyReveal}}($sid$) or \\(b) \emph{\texttt{PrivateKeyReveal}}($B$).
\end{enumerate}
\label{DEF:ECK-FRESH}
\end{Def}

In the eCK experiment the adversary is allowed to continue interacting with the parties even after issuing the test query with the restriction that the test session must remain fresh. 

\begin{Th}[eCK-security] A key agreement protocol is \emph{eCK-secure} if the following conditions hold:
\begin{enumerate}
\item If two honest parties complete matching sessions then, except with negligible probability, they both compute the same session key;
\item No polynomially bounded adversary can distinguish the session key of a fresh session from a randomly chosen session key, with probability greater than 1/2 plus a negligible function (in the security parameter).
\end{enumerate}
\label{TH:ECK-SEC}
\end{Th}

\subsection{Proof of eCK security of protocol IDAK}
\label{SEC:IDAKECK}

The following theorem proves that the two-pass IDAK key agreement protocol is secure in the eCK model under the Gap Bilinear Diffie Hellman and random oracle assumptions. 

\begin{Th}[IDAK eCK-security] If $\mathcal{H}, \mathcal{H}_1,\pi$ are random oracles, and the GBDH assumption holds in the group $G$ (from hereon $G$ will stand for $G_1$), then IDAK is an eCK-secure key agreement protocol.
Concretely, there exists a GBDH solver $S_D$ and a DLOG solver $S_L$ in $G$ such that 
\[
\texttt{\small{Adv}}_{\text{IDAK}}^{\text{eCK}}(E) \leq c_1\cdot \texttt{\small{Adv}}^{\text{DLOG}}(S_L)+ c_2\cdot\texttt{\small{Adv}}^{\text{GBDH}}(S_D) + \epsilon
\] 
where $\ell$ is the security parameter, $s(\ell)$ the number of activated sessions,  $N(\ell)$ the number of principals, $c_1=c_1(\ell)=s(\ell)^2$, $c_2=c_2(\ell)=s(\ell)N(\ell)$ and $\epsilon=\epsilon(\ell)=\emph{O}(s(\ell)^2/2^\ell)$. 
\label{TH:ECKE-1N-SEC}
\end{Th}

\emph{Proof}. Let $E$ denote a polynomially bounded adversary in the security parameter $\ell$. The adversary $E$ defeats protocol IDAK with non-negligible probability if it succeeds in the experiment described in Section \ref{SEC:ECK-MODEL} with probability $1/2 + \epsilon(\ell)$, where $\epsilon(\ell)$ is non-negligible. 
Guessing the answer of the Test query succeeds with probability $1/2$. Since $\mathcal{H}_1$ is a random oracle, $E$ can only distinguish a session key from a random string with probability significantly greater than $1/2$ if the adversary $E$ succeeds in one of the following attacks:
\begin{enumerate}
\item[A1:] (Replicate) $E$ forces a session to accept the same key of the test session even though the  two sessions are distinct and non-matching and obtains the secret key from the target session (by asking a \texttt{SessionKeyReveal} query);  
\item[A2:] (Forge) for two parties (say $A,B$) $E$ computes $\sigma\text{= }sk_{AB}\text{= }sk_{BA}$ and queries $\mathcal{H}_1$ with ($A,B,R_A,R_B,\sigma$) .
\end{enumerate}

For A1 to occur the adversary $E$ must query the random oracle $\mathcal{H}_1$ with the same 5-tuple used to compute the key of the test session. The attack succeeds if the random oracle produces collisions, since distinct sessions have distinct 5-tuples such collisions occur with negligible probability $\emph{O}(s(\ell)^2/2^\ell)$. 
Therefore, to succeed in the eCK experiment the adversary $E$ must perform a forging attack (A2). 
To this end, we show that if an adversary $E$ is able to win the eCK experiment then we may construct a GBDH solver in the underlying group $G$ which uses $E$ as a subroutine.
Algorithm $S$ simulates the eCK experiment in such a way that $E$'s view is indistinguishable from the real one. We consider the two cases described below.

\texttt{Case 1)} (matching sessions) The adversary $E$ is substantially passive (limiting itself only to eavesdropping the messages exchanged by any two parties) and chooses a test session with a matching session.  Let \texttt{match} denote the event that the adversary chooses either one of the matching sessions $sid$ or $sid^*$ owned respectively by $A,B$ as its test session (this event is conditioned on A2).
The description of algorithm $S$ follows:
\begin{enumerate}
	\item $S$ receives in input a GBDH challenge  ($g, X=g^x, Y=g^y, Z=g^z$) where $x,y,z \in_R Z_q^*$;
	\item $S$ establishes the identities ID$_i$ and the long-term keys ($d_{ID_i}, g_{ID_i}$) of $N(\ell)$ principals by calling algorithm \texttt{Extract};
	\item $S$ runs $E$ as a subroutine answering its queries as follows: a) \texttt{send} queries are simulated as usual, except for sessions $sid$ and $sid^*$ defined above; in this case assuming that ID$_i\equiv A$ and ID$_j\equiv B$ (for some $i,j$) $S$ selects the ephemeral keys $x,y \in_R Z_q^*$ for $A,B$ (respectively) and sets the messages $R_A=X, R_B=Y$. The session keys $sk_{AB}, sk_{AB}$ are set equal to a random value $G$ so $E$'s view of the protocol execution is the same as the real one because $sk_{AB}, sk_{AB}, X,Y \in G$ and $\mathcal{H}_1$ is a random oracle; b) all other queries are answered according to the protocol specification;
	\item When $E$ terminates, algorithm $S$ also terminates and outputs the same bit $b'$ as $E$ (as the input to the \texttt{Guess} query).  
\end{enumerate}
We claim that if $E$ succeeds in the eCK experiment then $S$ can solve the GBDH challenge. Indeed, if event \texttt{match} occurs, to succeed in the eCK experiment $E$ must query the random oracle $\mathcal{H}_1$ with the tuple $(A,B,R_A,R_B,\sigma$). To correctly compute $\sigma$, $E$ must obtain $x, y$ and to do so she can use any combination of queries that will maintain the test session fresh (recall that $E$ cannot reveal both $x,d_{ID_A}$ or $y,d_{ID_B}$). Therefore, $E$ must ask \texttt{PrivateKeyReveal} queries of both $A$ and $B$ (to obtain $d_{ID_A}$ and $d_{ID_B}$) and then the only way she can obtain $x$ (or $y$) is by computing $\emph{DLOG}_G(R_A)$ (or $\emph{DLOG}_G(R_B)$). As a result, the probability that $S$ succeeds is at least Pr[\texttt{match}$|$A2]/$s(\ell)^2$ since the simulation is perfect until event \texttt{match} occurs. This implies that $S$ can solve the discrete logarithm  in $G$ (i.e. algorithm $S_L$).     

\texttt{Case 2)} The adversary $E$ chooses a test session owned by $A$ that does not have a matching session; we denote this event by $\texttt{nomatch}$. $S$ receives in input a GBDH challenge ($g, X=g^x, Y=g^y, Z=g^z$) where $x,y,z \in_R Z_q^*$. To setup the simulation, $S$ establishes the identities ID$_i$ and the long-term keys ($d_{ID_i},g_{ID_i}$) of $N(\ell)-1$ honest principals except for a randomly chosen party $A$ for which she sets $d_A=X$. Now $S$ can correctly simulate all sessions (involving honest parties) invoked by $E$ during the eCK experiment except for those owned by $A$. When $E$ activates $A$, $S$ simulates $E$'s queries as follows:
\begin{enumerate}
	\item \texttt{Send} queries, $E$ activates a session $sid$ owned by $A$ and a peer $B$ who owns session $sid^{*}$. Regardless of $A$'s role in the communication  (i.e. as either initiator or responder), $S$ selects $\hat{x} \in_R Z_q^*$, runs \texttt{Send}($B,R_A=g_{ID_A}^{\hat{x}}$) and sets the session key of $A$ to $sk_{AB} \in_R G$. The value of $\hat{x}$ and $sk_{AB}$ are indistinguishable from the real values (in particular because $H_1$ is a random oracle). Note also that $S$ is perfectly aware that $B$ may be a fake party because she can handle any queries of the type \texttt{Extract}($B$) (for $B \not= $A$)$. 
	\item \texttt{EphemeralKeyReveal}($sid$) queries, for the $sid$ of a session owned by $A$ the answer is $\hat{x}$ (generated for $A$ by $S$, see above);
	\item \texttt{PrivateKeyReveal}($A$) queries, $S$ aborts; 
	\item \texttt{SessionKeyReveal}($sid$) queries, $S$ returns the key $sk_{AB}$ generated for $sid$ (see above); 
	\item queries of the random oracles and \texttt{Test} queries are simulated in the standard way.
\end{enumerate}

The simulator $S$ described above may fail either during the simulation of \texttt{Send} queries (or if $E$ asks the query \texttt{PrivateKeyReveal}($A$) as seen above). Indeed, assuming $A$ is the initiator (without loss of generality) and event $\texttt{nomatch}$ has occurred, $E$ can distinguish between the real and simulated worlds if she computes the session key $sk_{BA}$ by querying $\mathcal{H}_1$  with the tuple ($A,B,R_A, R_B=g_{ID_B}^{\hat{y}}, \sigma$) where $\sigma=\hat{e}(g_{ID_A}^{s_A}\cdot R_A, d_{ID_B}^{(\hat{y}+s_B)})$ and $\hat{y} \in_R Z_q^*$; by choosing $sid$ as the test session and asking the query \texttt{SessionKeyReveal}($sid$) $E$ is able to determine that $sk_{AB} \ne sk_{BA}$ (this implies that session $sid$ is not fresh and thus $E$ cannot choose it as the test session). However, since event A2 has occurred, $S$ can intercept $E$'s queries of the random oracle  $\mathcal{H}_1$ having as arguments the tuple ($A,B,R_A,R_B,\sigma$) and respond with $sk_{AB}$ if DBDH($(X, g_{ID_B}^{\hat{x}+s_A},g_{ID_B}^{\hat{y}+s_B},\sigma$)=1. Therefore, the only way $E$ can detect that it is running in a simulated experiment is by asking \texttt{PrivateKeyReveal} queries of both $A$ and $B$ (to obtain $\hat{x}$ and $\hat{y}$) and $DLOG_G(X)$; this event occurs with probability at least Pr[\texttt{nomatch}$|$A2]/$s(\ell)N(\ell)$ thus allowing $S$ to solve the GBDH  problem in $G$ (i.e. algorithm $S_D$). 

We note that in case $A$ is the responder then $E$ can choose $sid$ as the test session but cannot ask  a  \texttt{PrivateKeyReveal}($A$) query. Therefore again to detect the simulation $E$ must ask a query of the type $DLOG_G(X))$.

\section{Conclusions}

The IDAK protocol exhibits an excellent design, interesting security features and  high computational  efficiency, which make it a good candidate for resource constrained devices. 

In this report we have proven that the IDAK protocol is secure in the extended Canetti-Krawczyk model, the later is stronger than the Bellare-Rogaway model since it intrinsically accounts for KCI attacks and other types of possible threats.

As the result, the protocol can be used to secure communications between peers in insecure networks (e.g. the Internet) that are totally under control of powerful active adversaries.
   
\bibliographystyle{abbrv}

\end{document}